\documentclass[aps,prl,superscriptaddress,twoside,twocolumn,
nofootinbib,10pt,showpacs,floatfix]{revtex4-1}
\usepackage{graphicx}
\usepackage{mathrsfs}
\usepackage{amsmath,bbm}
\usepackage{CJK}
\usepackage[colorlinks,hyperindex,citecolor=black,
	linkcolor=black,urlcolor=black]{hyperref}
\newcommand\tr{\mathop{\mathrm{Tr}}}

\begin{document}

\begin{CJK*}{UTF8}{gbsn}
\title{Magnetic response of baryon properties in a skyrmion model}
\author{Bing-Ran He (何秉然)}
\email[E-mail: ]{he@hken.phys.nagoya-u.ac.jp}
\affiliation{
Department of Physics, Nagoya University, Nagoya 464-8602, Japan
}

\date{\today}

\begin{abstract}
An axially symmetric ansatz is proposed to investigate the properties of baryon in a uniform magnetic field. The baryon number is shown to be conserved, while the baryon shape is stretched along the magnetic field. It is found that with increasing magnetic field strength, the static mass of the baryon first decreases and then increases, while the size of the baryon first increases and then decreases. Finally, in the core part of the magnetar, the equation of state strongly depends on the magnetic field, which modifies the mass limit of the magnetar.
\end{abstract}

%12.39.Dc Skyrmions
%12.39.Fe Chiral Lagrangians
%11.30.Rd Chiral symmetries


\maketitle
\end{CJK*}
%\section{Introduction}
%%%%%%%%%%%%%%%%%%%%%%%%%%%%%%%%%%%%%%%%%%%%%%%%%%%%%%%%%%%%%%%%%%%%%%%%%%%%%%%%%%%%%%%%%%%%%%%%%%
\emph{Introduction.---}
The behavior of hadrons in an ultrastrong magnetic field is very rich and subtle. On one hand, recent experiments have observed that a strong magnetic field exists in non-central heavy-ion collisions. 
On the other hand, an ultrastrong magnetic field widely exists in the early universe and in magnetars. Thus, a number of studies have been performed in the last few years~\cite{Kharzeev_Miransky}. 
Generally, in an ultrastrong magnetic field background, hadrons have two important features: 
(i) the $SO(3)_{\rm space}$ and $SU(2)_{\rm flavour}$ symmetry of the hadron is explicitly broken by the magnetic field, and (ii) the hadron is strongly deformed in a strong magnetic field. 
Recently, several models have been developed that predict the mass and shape change of the meson in a uniform magnetic field. For example, lattice quantum chromodynamics (QCD) calculations imply the quark potential changes in the magnetic field background~\cite{Bonati:2014ksa}; however, because it is difficult to predict the charge radii of baryons in a vacuum using lattice QCD calculations~\cite{Abdel-Rehim:2015jna}, the shape of the baryon in a strong magnetic field background is still unclear. 

In this letter, the focus is to study the baryon mass and shape under the confinement phase in a strong magnetic field background. 
Because the pion does not condense in a strong magnetic field, Skyrme showed that the properties of a baryon could be evaluated by constructing a soliton object in mesonic models~\cite{Skyrme:1962vh}. 
The skyrmion preserves both energy and shape, which helps in understanding the baryon properties in strong magnetic field backgrounds. 
In the present study, the baryon number is shown to be always conserved, as the presence of the magnetic field does not break the $U(1)_V$ symmetry. 
It is found that the shape of the baryon is stretched along the magnetic field, considered as the z-axis, as the symmetry of $SO(3)_{\rm space}$ is explicitly broken to $SO(2)_{\rm space}$ in a strong magnetic field background. 
Furthermore, with increasing magnetic field strength, the dominant role shifts from the linear term of the magnetic field to the quadratic term of the magnetic field; consequently, the static mass of the baryon first decreases and then increases. 
Finally, because the mass and shape of the baryon are both influenced by the magnetic field, the equation of state (EOS) for magnetars is modified, which implies that effects beyond the mean-field approximation should be considered to predict the mass limit of magnetars correctly. 

%\section{The model}
%%%%%%%%%%%%%%%%%%%%%%%%%%%%%%%%%%%%%%%%%%%%%%%%%%%%%%%%%%%%%%%%%%%%%%%%%%%%%%%%%%%%%%%%%%%%%%%%%%
\emph{The model.---}
The model contains two parts:
\begin{equation}
\Gamma=\int d^4 x \mathscr{L}+\Gamma_{\rm WZW}\,,
\label{total_action}
\end{equation}
where the first term is the even intrinsic parity and second term is the odd intrinsic parity.

The even intrinsic parity part is expressed as
\begin{eqnarray}
\mathscr{L}&=&\frac{f_\pi^2}{16}\tr(D_\mu U ^\dag D^\mu U) + \frac{1}{32g^2}\tr([U ^\dag D_\mu U, U^\dag D_\nu U]^2)\nonumber\\
&&+\frac{m_\pi^2 f_\pi^2}{16}\tr(U+U^\dag-2)
\,,\label{lagrangian}
\end{eqnarray}
where $f_\pi$ is the pion decay constant, $m_\pi$ is the pion mass, and $g$ is a dimensionless coupling constant. 
The covariant derivative for $U$ is $D_\mu U = \partial_\mu U - i \mathcal{L}_\mu U + i U \mathcal{R}_\mu$,  
where $\mathcal{L}$ and $\mathcal{R}$ are the external fields expressed as  $\mathcal{L}_\mu=\mathcal{R}_\mu= e Q_{\rm B} {\mathcal{V}_B}_\mu + e Q_{\rm E} H_\mu$ for the present purpose, 
$e$ is the unit electric charge, $Q_{\rm B}=\frac{1}{3}\mathbbm{1}$ is the baryon number charge matrix, $Q_{\rm E}=\frac{1}{6}\mathbbm{1}+\frac{1}{2}\tau_3$ is the electric charge matrix,  $\mathbbm{1}$ is the rank $2$ unit matrix, $\tau_3$ is the third Pauli matrix, and ${\mathcal{V}_B}_\mu$ is the external gauge field of the $U(1)_{V}$ baryon number. The magnetic field $H_\mu$ is expressed as 
$H_\mu  = - \frac{1}{2} B  y g_\mu^{\;\;1} + \frac{1}{2} B  x g_\mu^{\;\;2}$ 
in the symmetric gauge. 

The odd intrinsic parity part, i.e., Wess--Zumino--Witten (WZW) action $\Gamma_{\rm WZW}\equiv\int d^4x \mathscr{L}_{\rm WZW}$, represents the chiral anomaly effects; the odd intrinsic parity is given in Ref.~\cite{Brihaye:1984hp}.

%\section{The ansatz}
%%%%%%%%%%%%%%%%%%%%%%%%%%%%%%%%%%%%%%%%%%%%%%%%%%%%%%%%%%%%%%%%%%%%%%%%%%%%%%%%%%%%%%%%%%%%%%%%%%
\emph{The ansatz.---}
For the present study, 
when magnetic field is small, the ansatz is the spherically symmetric hedgehog ansatz with  $SO(3)_{\rm spin}\times SU(2)_{\rm isospin}$ symmetry, which is widely used in other studies; when the magnetic field is strong, the space and isospin symmetry are explicitly broken by the magnetic field, and the ansatz has the axially symmetric $SO(2)_{\rm spin}\times U(1)_{\rm isospin}$ symmetry. 
Furthermore, when $e B\neq0$, the directions of the unbroken symmetries are correlated with each other, i.e., the third direction of the isospin space, $\tau_3$, is always along the direction of magnetic field, namely the $z$-axis. 
With respect to the symmetry, $U$ is decomposed as 
\begin{eqnarray}
U=\cos(F(r))\mathbbm{1} + \frac{i \sin(F(r))}{r}\Big(  \frac{\tau_1}{c_\rho}x + \frac{\tau_2}{c_\rho}y +  \frac{\tau_3}{c_z}z\Big)\,,\label{uansatzs}
\end{eqnarray}
where
\begin{eqnarray}
x&=&c_\rho r \sin (\theta )\cos (\varphi ) \,,\nonumber\\
y&=&c_\rho r \sin (\theta )\sin (\varphi ) \,,\nonumber\\
z&=&c_z r \cos (\theta )\,.\label{xyzansatzs}
\end{eqnarray}
Here, $c_\rho$ and $c_z$ are positive dimensionless parameters, $\rho=\sqrt{x^2+y^2}$, and $\theta$ and $\varphi$ are polar angles with $\theta\in[0,\pi]$ and $\varphi\in[0,2\pi]$. 
A similar ansatz has been previously described \cite{Holzwarth:1985rb} that only deforms the spatial space; however, in this study, a further assumption is made that both the spatial space and isospin space are deformed. 

%\section{The baryon number}
%%%%%%%%%%%%%%%%%%%%%%%%%%%%%%%%%%%%%%%%%%%%%%%%%%%%%%%%%%%%%%%%%%%%%%%%%%%%%%%%%%%%%%%%%%%%%%%%%%
\emph{The baryon number.---}
The baryon number current of the model is obtained by a functional derivative of the WZW term 
with ${\mathcal{V}_B}_\mu$, i.e., $j_B^\mu=\frac{\partial\mathscr{L}_{\rm WZW}}{\partial(e{\mathcal{V}_B}_\mu)}|_{{\mathcal{V}_B}_\mu\to0}$.
Combined with \eqref{uansatzs} and \eqref{xyzansatzs}, the time component of the baryon current is obtained as
\begin{eqnarray}
j_B^0(r,\theta)&=&-\frac{1}{c_\rho^2 c_z}\frac{F' \sin ^2(F)}{2 \pi ^2 r^2}
+ e B\frac{ \sin ^2(\theta ) \sin (F) \cos (F)}{8 \pi ^2 r c_z}\nonumber\\
&&+e B\frac{  F' \left(\cos ^2(\theta )-\sin ^2(\theta ) \sin ^2(F)\right)}{8 \pi ^2  c_z}\,.
\label{baryon_number_current}
\end{eqnarray}
The baryon number $N_B$ is obtained as $N_B=\int dV j_B^0(r,\theta)=\frac{\sin (2 F) \left(e B c_{\rho }^2 r^2 +6\right)-12 F}{12 \pi }\Big|^{F(\infty)=0}_{F(0)=\pi} =1$, 
where $dV=c_\rho^2 c_z r^2 \sin (\theta ) dr d\theta d\varphi$. In the present calculation, the boundary conditions 
 $F(0)=\pi$ and $F(\infty)=0$ are imposed.  
Here, the baryon number $N_B$ is shown to be \emph{always} normalized as one, which consistent with the fact that the presence of the magnetic field does not break the $U(1)_V$ symmetry. 
In Ref.~\cite{Eto:2011}, the baryon number is not conserved when $e B\neq0$; however, this result is incorrect because it is based on the expression of baryon current shown in Ref.~\cite{Brihaye:1984hp}, which is incorrect. Thus, Ref.~\cite{Eto:2011} excludes the last term in the second line of Eq.~\eqref{baryon_number_current}.

%\section{Numerical results }
%%%%%%%%%%%%%%%%%%%%%%%%%%%%%%%%%%%%%%%%%%%%%%%%%%%%%%%%%%%%%%%%%%%%%%%%%%%%%%%%%%%%%%%%%%%%%%%%%%
\emph{Numerical results.---} 
Let us consider parameters $c_\rho$ and $c_z$. The combination of these parameters causes two effects: (i) it twists the shape, and 
(ii) it scales the volume. 
To separate these two effects, $c_{\rho,z}$ is rewritten as $c_{\rho,z}\equiv \lambda \bar{c}_{\rho,z}$, 
where $\lambda$ is a parameter representing the scale effect, and $\bar{c}_{\rho,z}$ are the parameters represent the twist effect, which satisfies $\bar{c}_\rho^2 \bar{c}_z\equiv 1$. 
Because the scale effect of $c_{\rho,z}$ can be absorbed by performing the 
scale transform $r\to  r/\lambda$ with no loss of generality, 
we address $c_{\rho,z}\equiv \bar{c}_{\rho,z}$, where $c_\rho\equiv 1/\sqrt{c_z}$.

Substituting the ansatz equations \eqref{uansatzs} and \eqref{xyzansatzs} in \eqref{total_action}, the soliton mass 
$M_{\rm sol}=\int dV \mathscr{M}_{\rm sol}(r,\theta)$ is obtained. 
Following Ref.~\cite{Adkins:1983ya}, after properly choosing the Skyrme units, a standard set of parameters is considered: 
$m_\pi=138$ [MeV], $f_\pi=108$ [MeV], and $g=4.84$. 

\begin{figure}[htb]
\centering
\includegraphics[scale=0.42]{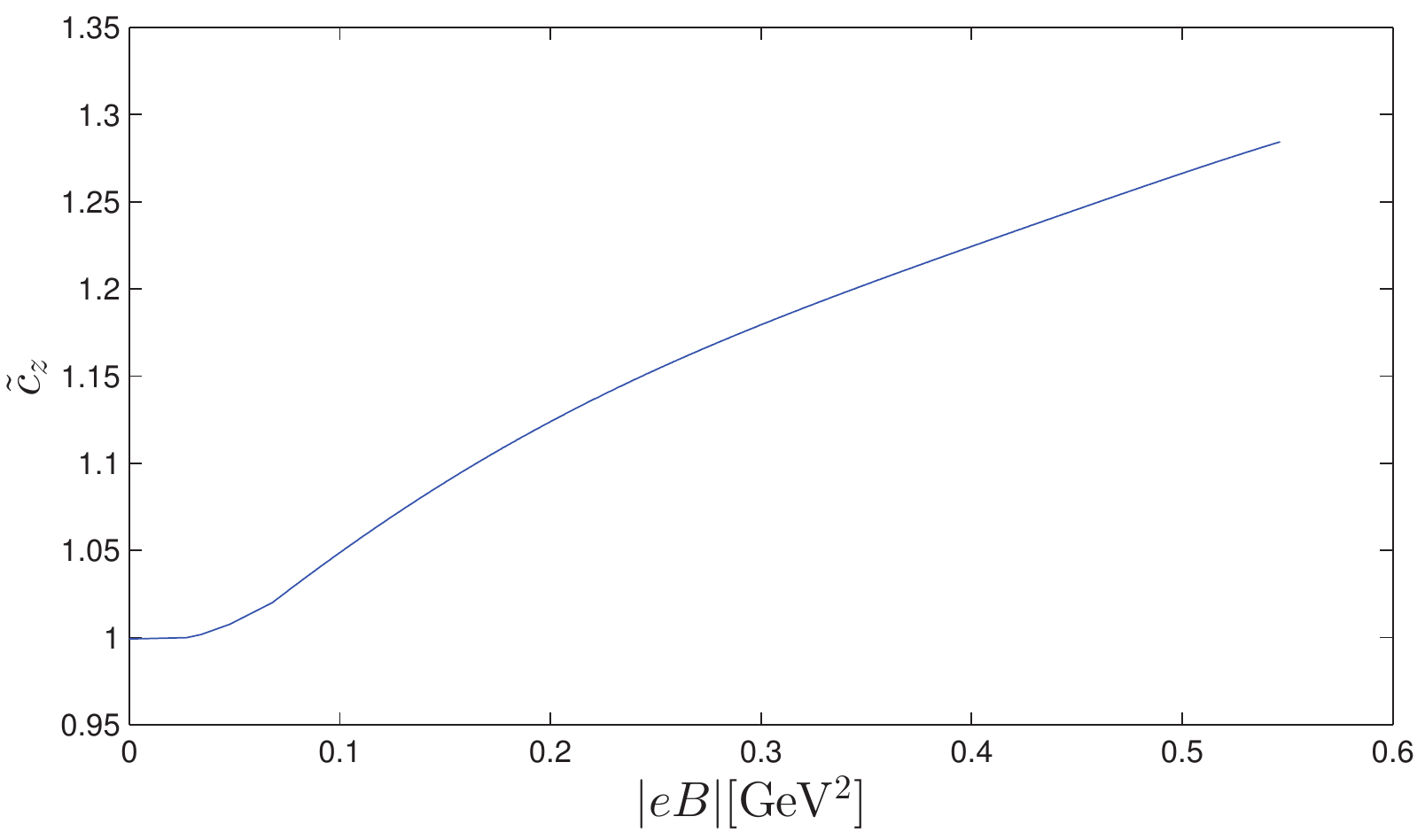} 
\caption{$|e B|$ dependence of $\tilde{c}_z$.
}
\label{fig:bmag_cz}
\end{figure}

The soliton mass $M_{\rm sol}$ depends on two parameters: the strength of the magnetic field $|e B|$ and the parameter $c_z$. 
Here, the value of $c_z$ that minimizes the soliton mass for a given $|e B|$ is $\tilde{c}_z$. 
The $|e B|$ dependence of $\tilde{c}_z$ is shown in Fig.~\ref{fig:bmag_cz}. 
%%%figure%%%%%%%%%%%%%%%%%%%%%%%%%%%%%%%
When $|eB|<0.027~{\rm [GeV^2]}$, the ansatz of the skyrmion is spherically symmetric ($\tilde{c}_z=1$); however, when $|eB|>0.027~{\rm [GeV^2]}$, the ansatz of the skyrmion becomes axially symmetric ($\tilde{c}_z>1$), which implies the baryon is stretched along the magnetic field, the z-axis.

\begin{figure}[htb]
\centering
\includegraphics[scale=0.42]{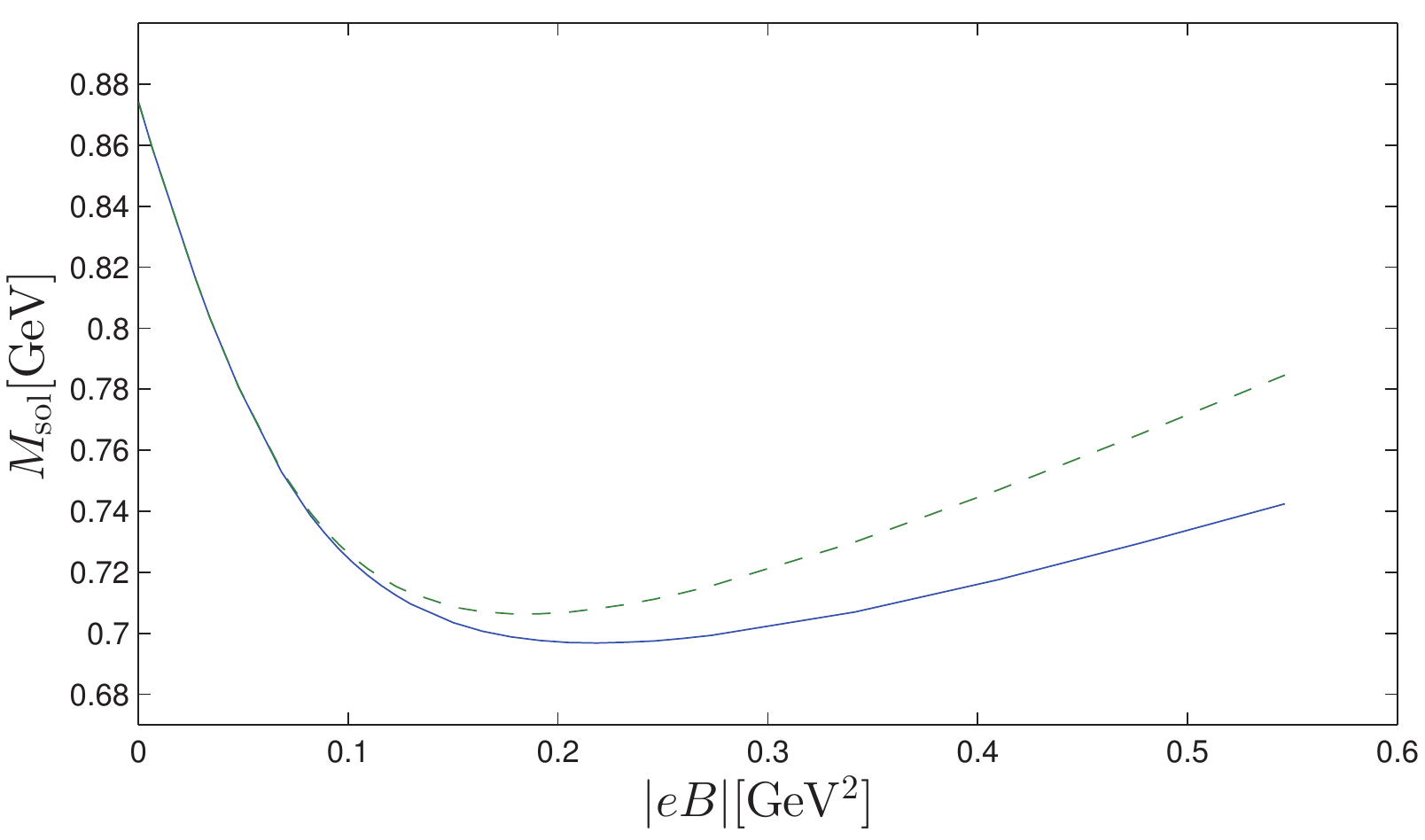} 
\caption{$|e B|$ dependence of soliton mass ${M}_{\rm sol}$. The green dashed line represents $\bar{M}_{\rm sol}$ ($c_z\equiv 1$), and the blue curve represents $\tilde{M}_{\rm sol}$ ($c_z\equiv \tilde{c}_z$). 
}
\label{fig:bmag_msol}
\end{figure}

To compare the difference between the spherically symmetric hedgehog ansatz ($c_z\equiv 1$) and the newly proposed ones of \eqref{uansatzs} and \eqref{xyzansatzs} ($c_z\equiv \tilde{c}_z$), the corresponding soliton mass for a fixed $|e B|$ is addressed as $\bar{M}_{\rm sol}$ and $\tilde{M}_{\rm sol}$, respectively. 
The $|e B|$ dependence of $\bar{M}_{\rm sol}$ and $\tilde{M}_{\rm sol}$ is shown in Fig.~\ref{fig:bmag_msol}.
%%%figure%%%%%%%%%%%%%%%%%%%%%%%%%%%%%%%
Since $\tilde{M}_{\rm sol}\le\bar{M}_{\rm sol}$, the physical mass of the baryon is $\tilde{M}_{\rm sol}$. 
The baryon mass $\tilde{M}_{\rm sol}$ first decreases and then increases as the strength of the magnetic field increases because the dominant role shifts from the term proportional to $(eB)$ to the term proportional to $(eB)^2$ in $M_{\rm sol}$~\cite{Gladikowski_Schroers}. 

In the following discussion, $c_z\equiv \tilde{c}_z$ is considered to stabilize the skyrmion. 
To evaluate the twist effect of the baryon caused by the magnetic field, the root-mean-square (RMS) radius for the baryon number density and energy density is defined as $\langle X^2 \rangle_{B}^{1/2}=\sqrt{\int_0^\infty dV X^2 j_{B}^0 }$ and $\langle X^2 \rangle_{E}^{1/2}=\sqrt{\frac{1}{M_{\rm sol}}\int_0^\infty dV X^2 \mathscr{M}_{\rm sol} }$, 
where $X$ represents $R_x$, $R_z$, and $R$. Here, $R \equiv \sqrt{x^2+y^2+z^2}$, and $R_x$ and $R_z$ represent the projection of $R$ on the $x$ and $z$ axis, respectively.

\begin{figure}[htb]
\centering
\includegraphics[scale=0.42]{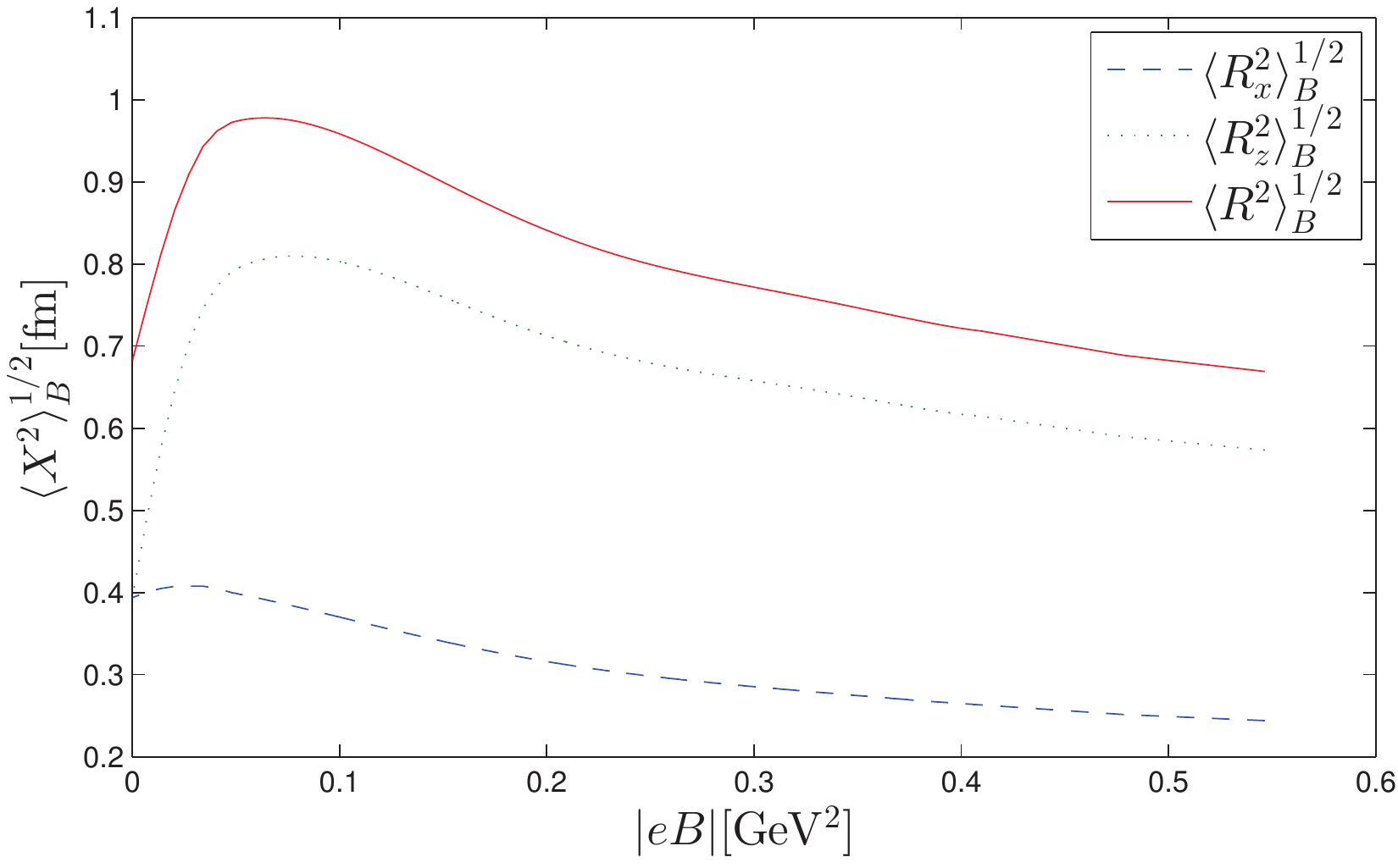} 
\caption{$|e B|$ dependence of the RMS radius for the baryon number density.
}
\label{fig:bmag_RMS_B}
\end{figure}

\begin{figure}[htb]
\centering
\includegraphics[scale=0.42]{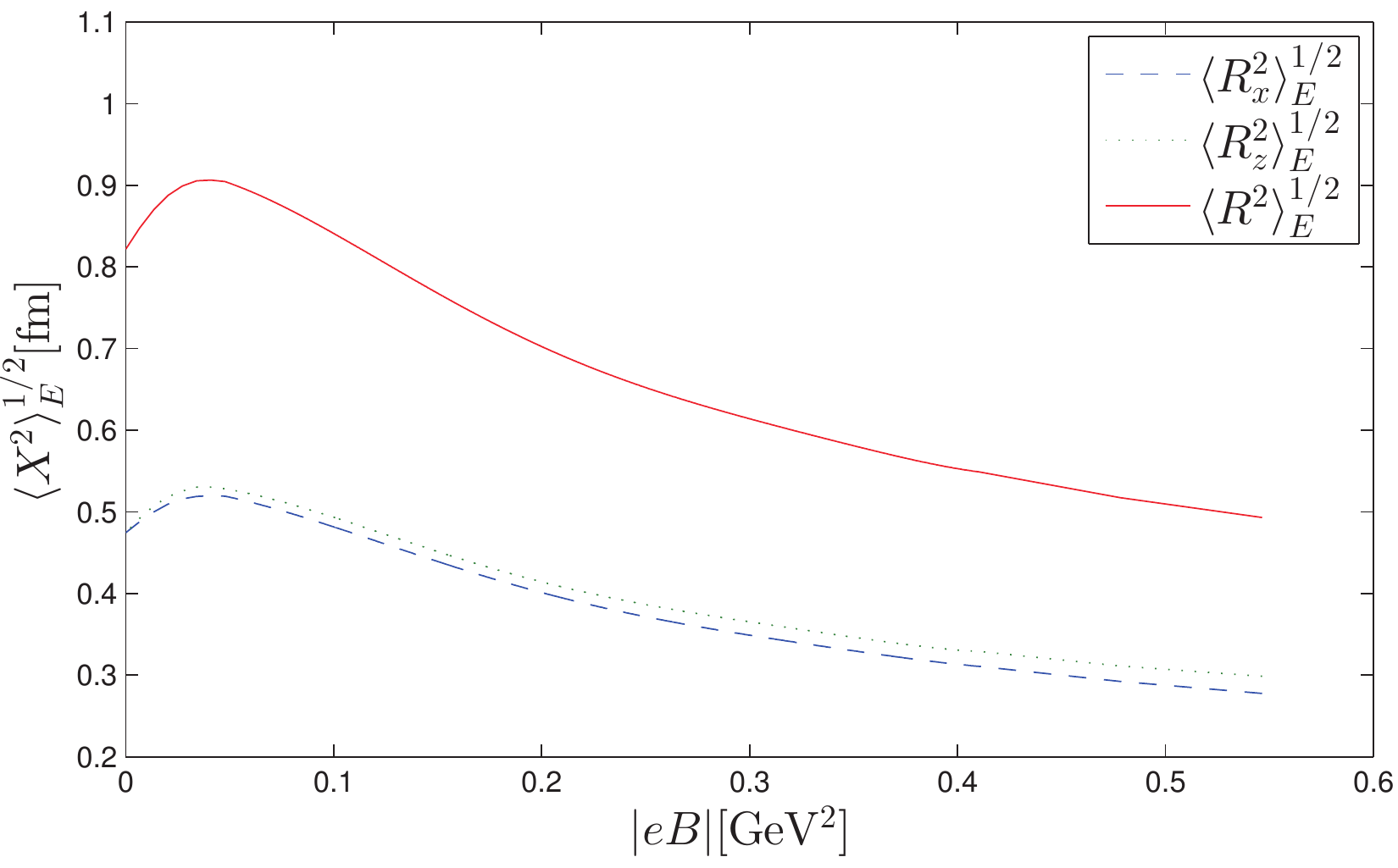} 
\caption{$|e B|$ dependence of the RMS radius for the energy density.
}
\label{fig:bmag_RMS_E}
\end{figure}

\begin{figure}[htb]
\centering
\includegraphics[scale=0.42]{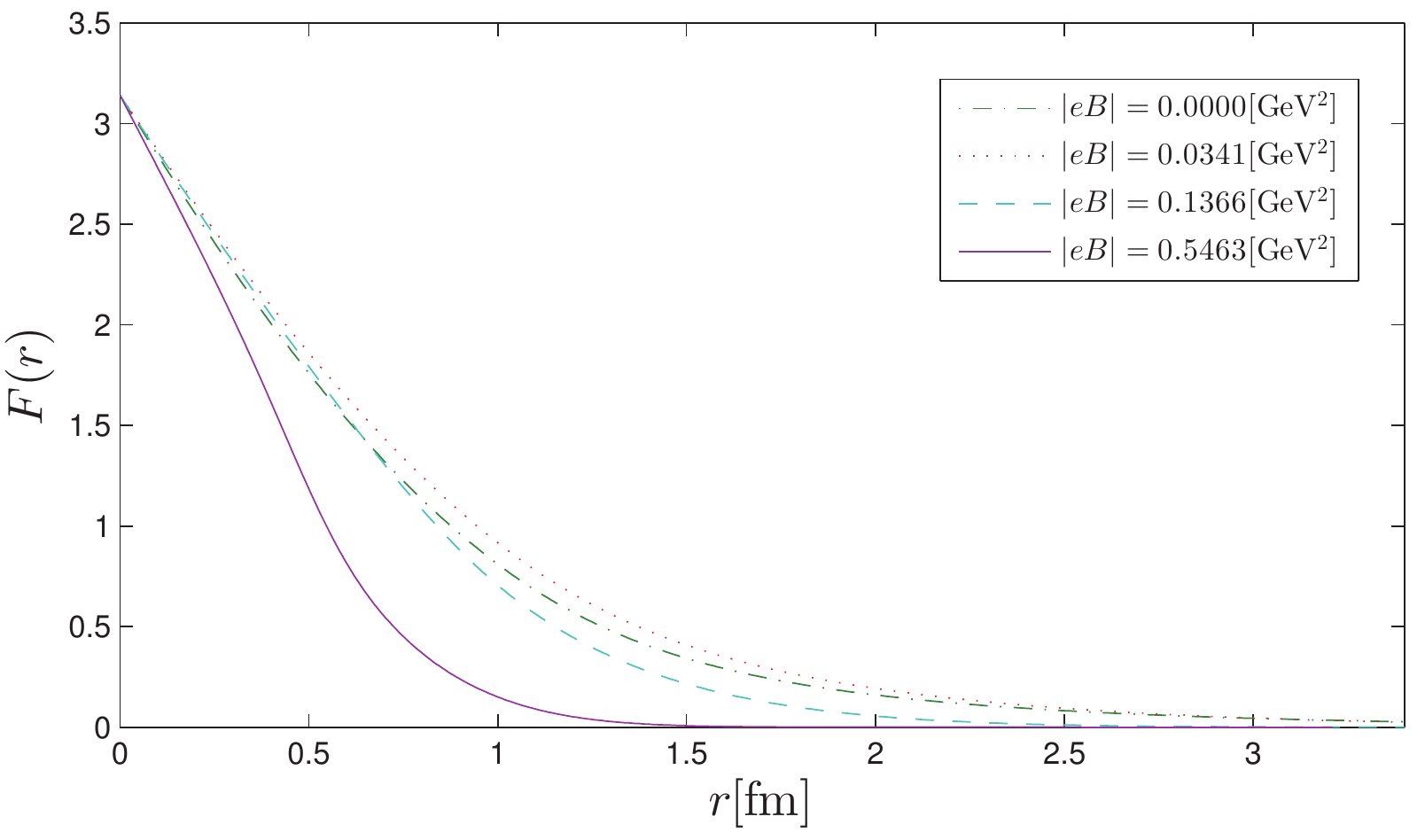} 
\caption{Profile functions 
 of the stabilized skyrmion, when $|e B| =0, 0.0341, 0.1366,$ and $ 0.5463 ~[{\rm GeV}^2]$.
}
\label{fig:eb_f_profile}
\end{figure}

The RMS radii for the baryon number density and energy density are shown in Fig.~\ref{fig:bmag_RMS_B} and Fig.~\ref{fig:bmag_RMS_E}, respectively.
%%%figure%%%%%%%%%%%%%%%%%%%%%%%%%%%%%%%
%%%figure%%%%%%%%%%%%%%%%%%%%%%%%%%%%%%%
Fig.~\ref{fig:bmag_RMS_B} and Fig.~\ref{fig:bmag_RMS_E} show that 
when $|eB|$ is weak, the RMS radii increase; on the other hand, when $|eB|$ is strong, the RMS radii decrease. 
This tendency can be understand from the profile functions of $F(r)$, which are shown in Fig.~\ref{fig:eb_f_profile}.
%%%figure%%%%%%%%%%%%%%%%%%%%%%%%%%%%%%%
Fig.~\ref{fig:eb_f_profile} shows that for weak $|eB|$ ($|eB|=0.0341~[{\rm GeV}^2]$), the increase of $|eB|$ broadens the profile function. On the other hand, for strong $|eB|$ ($|eB|= 0.5463 ~[{\rm GeV}^2]$), the increase of $|eB|$ narrows the profile function. 
There is another explanation. For weak $|e B|$, the baryon mass decreases, which causes the baryon size to increase, and the RMS radii will increase. For strong $|e B|$, the freedom of the charged meson ($\pi^{+,-}$) is restricted in the x-y plane, which causes the RMS radii to decrease.
Therefore, the RMS radii first increase and then decrease as $|eB|$ increases. 

%%%figure%%%%%%%%%%%%%%%%%%%%%%%%%%%%%%%
\begin{figure}[htb]
\centering
\includegraphics[scale=0.56]{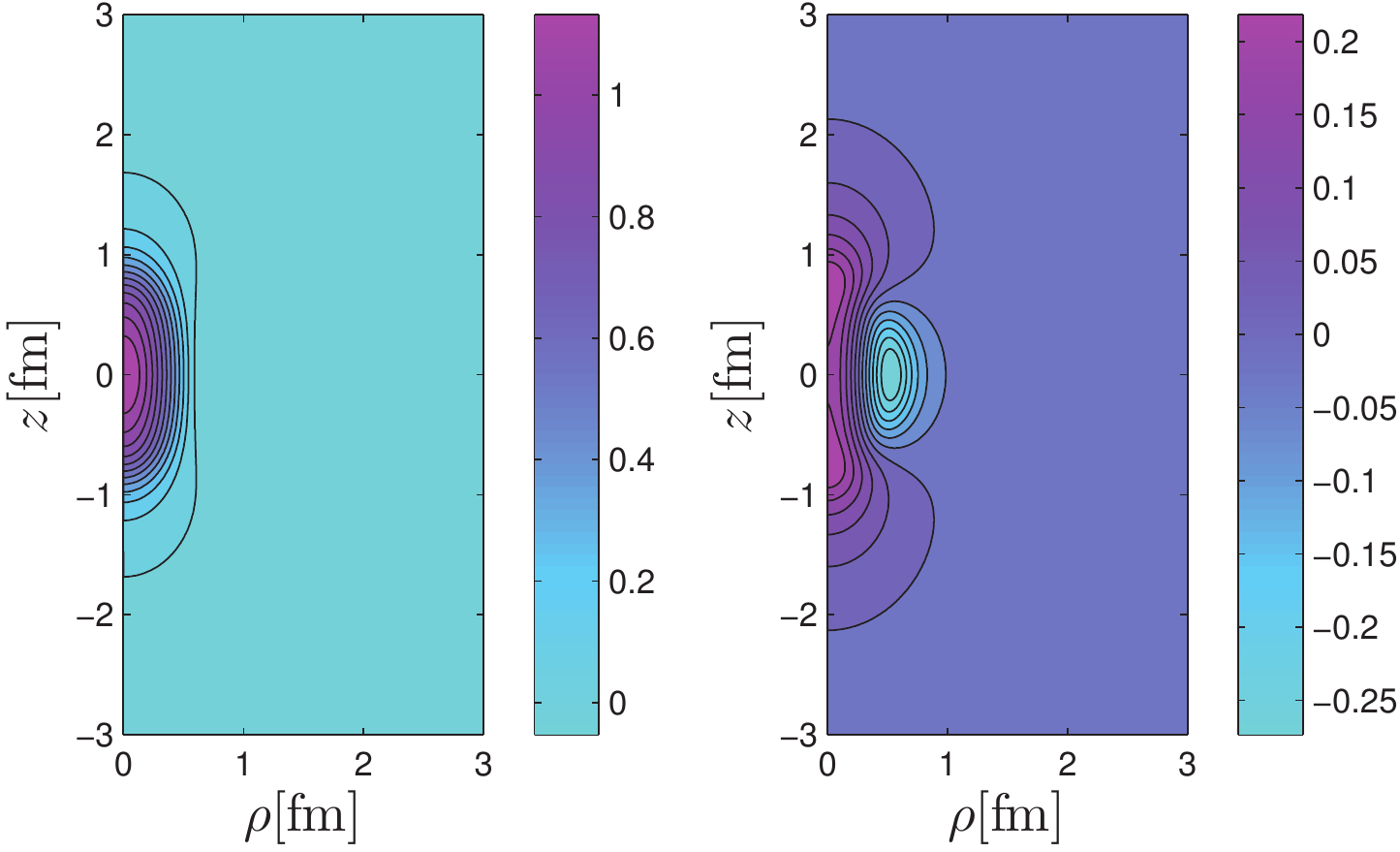}
\caption{The baryon number density contour plot when $|e B| =0.5463 ~[{\rm GeV}^2]$. The left figure is the total baryon number density, and the right figure is the induced baryon number density.
}
\label{fig:profile_baryon_charge}
\end{figure}
%%%figure%%%%%%%%%%%%%%%%%%%%%%%%%%%%%%%
\begin{figure}[htb]
\centering
\includegraphics[scale=0.56]{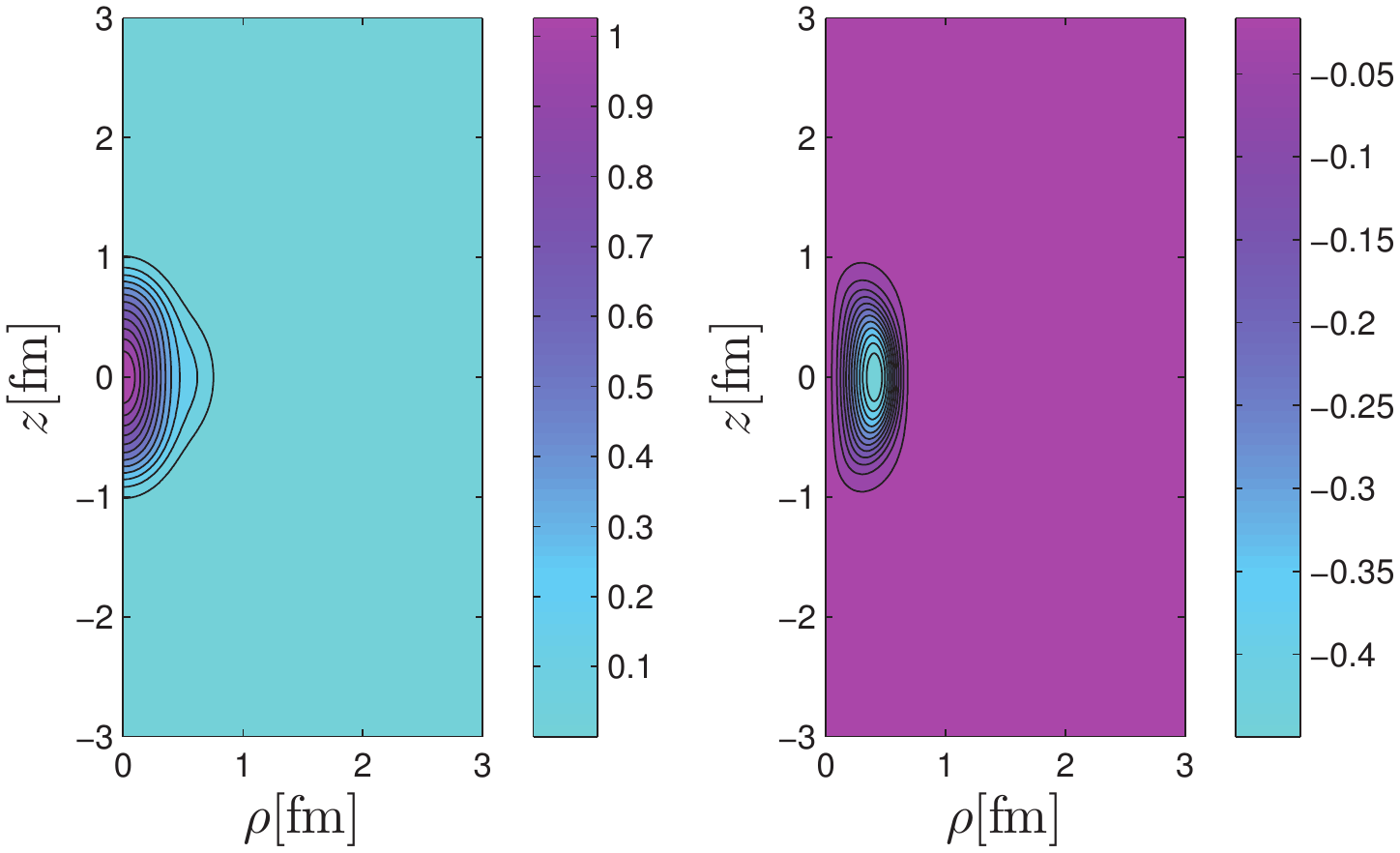} 
\caption{The energy density contour plot when $|e B| =0.5463 ~[{\rm GeV}^2]$. The left figure is the total energy density, and the right figure is the induced energy density.
}
\label{fig:profile_energy}
\end{figure}

%\section{Anisotropic of baryon number density and energy density}
%%%%%%%%%%%%%%%%%%%%%%%%%%%%%%%%%%%%%%%%%%%%%%%%%%%%%%%%%%%%%%%%%%%%%%%%%%%%%%%%%%%%%%%%%%%%%%%%%%
\emph{Anisotropic of baryon number density and energy density.---}  
Fig.~\ref{fig:bmag_RMS_B} shows that $\langle R_z^2 \rangle_{B}^{1/2}$ is larger than $\langle R_x^2 \rangle_{B}^{1/2}$. This behavior is understandable from the baryon number density contour plot in Fig.~\ref{fig:profile_baryon_charge}.  
Fig.~\ref{fig:profile_baryon_charge} shows that 
when $|e B|=0.5463 ~[{\rm GeV}^2]$, the shape of the total baryon number density is stretched along the z-axis, but the induced baryon number density has both positive and negative distributions. The positive part is found along the z-axis, while the negative part is away from z-axis but gathering around the x-y plane. 
Because of the total baryon number density along the z-axis, $\langle R_z^2 \rangle_{B}^{1/2}$ is larger than $\langle R_x^2 \rangle_{B}^{1/2}$. 

Fig.~\ref{fig:bmag_RMS_E} shows that $\langle R_z^2 \rangle_{E}^{1/2}$ is slightly larger than $\langle R_x^2 \rangle_{E}^{1/2}$; however, this mechanism is quite complicated.  
The energy density contour plot shown in Fig.~\ref{fig:profile_energy} shows that 
when $|e B|=0.5463 ~[{\rm GeV}^2]$, the total energy density has two parts: the high energy density part found along the z-axis and the low energy density part away from the z-axis. 
The induced energy density is stretched along the z-axis and contributes to both the z-axis and x-y plane. 
Because the induced energy always contributes to both the z-axis and x-y plane, $\langle R_z^2 \rangle_{E}^{1/2}$ is slightly larger than $\langle R_x^2 \rangle_{E}^{1/2}$.

%\section{Conclusions and discussions }
%%%%%%%%%%%%%%%%%%%%%%%%%%%%%%%%%%%%%%%%%%%%%%%%%%%%%%%%%%%%%%%%%%%%%%%%%%%%%%%%%%%%%%%%%%%%%%%%%%
\emph{Conclusions and discussions.---}
In this letter, the properties of baryon in a uniform magnetic field were studied. An axially symmetric ansatz was employed to investigate the shape deformation of the baryon caused by the magnetic field. 

It was first shown that the baryon number is always correctly normalized as one, while the shape of baryon is stretched along the magnetic field, the z-axis, for a nonzero magnetic field.

Next, the static mass of the baryon was investigated. With an increase of the magnetic field strength $|eB|$, the baryon mass first decreases then increases because the dominant role shifts from the $(eB)$ term to the $(eB)^2$ term in $M_{\rm sol}$. 
When $|eB|\sim 11 m_\pi^2 $, the baryon mass has a minimal point, which decreases the mass by about $20\%$ compared to the mass in vacuum.

The RMS radii of the baryon number density and the energy density were also investigated. 
It was found that the RMS radii of $\langle R^2 \rangle_{B}^{1/2}$ and $\langle R^2 \rangle_{E}^{1/2}$ first increase and then decrease with the increase of the magnetic field. When $|eB|\sim 3 m_\pi^2 $, $\langle R^2 \rangle_{B}^{1/2}$ has a maximum point that increases by about $43\%$ over that in vacuum. On the other hand, when $|eB|\sim 2 m_\pi^2 $, $\langle R^2 \rangle_{E}^{1/2}$ has a maximum point that increases by about $10\%$ over that in vacuum.  

Further, it was found that the distribution of the baryon number density and energy density is anisotropic. 
For the anisotropy of $\langle R^2 \rangle_{B}^{1/2}$ and $\langle R^2 \rangle_{E}^{1/2}$, it was found that the baryon number density strongly depends on the increase of the magnetic field, which causes  $\langle R_z^2 \rangle_{B}^{1/2}$ to be larger than $\langle R_x^2 \rangle_{B}^{1/2}$. For example, when $|eB|\sim 2 m_\pi^2 $, the ratio of $\langle R_z^2 \rangle_{B}^{1/2}/\langle R_x^2 \rangle_{B}^{1/2}$ is about $1.9$. The energy density does not strongly depend on the increase of the magnetic field, which causes $\langle R_z^2 \rangle_{E}^{1/2}$ to be slightly larger than $\langle R_x^2 \rangle_{E}^{1/2}$. 
In addition, it was found that $\langle R_z^2 \rangle_{B,E}^{1/2}>\langle R_x^2 \rangle_{B,E}^{1/2}$ is satisfied for all studied ranges of magnetic field strength. This result is consistent with the fact that the baryon is stretched along the magnetic field, the z-axis.

For a strong magnetic field ($|eB|> m_\pi^2$), the main behaviors of the baryon that were observed range 
 from $2 m_\pi^2$ to $11 m_\pi^2$, which is lower than the strength of magnetic field that can be obtained in the Large Hadron Collider ($|e B| \sim 15 m_\pi^2$)~\cite{Skokov:2009qp}. 
It is expected that the dependence of mass and RMS radii on magnetic field could be confirmed using future non-central collider experimental observations.

For a weak magnetic field ($|eB|< m_\pi^2$) in the core part of the magnetar ($|eB|\sim 10^{-2}~[{\rm GeV}^2]$)~\cite{Harding:2006qn}, the baryon mass decreases  
and the radii increase. 
Therefore, the mass-radius relation for magnetars is modified. A simple estimation shows that the baryon density decreases by about $28\%$ and the energy density decreases by about $14\%$ compared to that in vacuum, which significantly modifies the EOS. 
In many astrophysics applications, the baryon is used as a mean-field approximation for predicting the mass limit of compact stars; however, the present letter shows that effects beyond the mean-field theory, i.e., the influence of a magnetic field, should be considered for predicting the mass limit of magnetars correctly. 
For example, the skyrmion crystal approach and/or BPS skyrmion approach could provide more precise predictions of the effects of a magnetic field on EOS for magnetars~\cite{Dong_Wereszczynski}. 

In the present analysis, the RMS radii of the baryon number density and energy density have some model dependence; however, since it is difficult to predict the RMS radii of baryons, even in vacuum, by using lattice QCD \cite{Abdel-Rehim:2015jna}, the present study shows important implications of the effects of a magnetic field on baryons for both weak and strong magnetic fields~\cite{Adam:2014xfa}. 
For an ultrastrong magnetic field ($|eB|> m_{\rho,\omega}^2$), the behaviors of vector and scalar mesons also play important roles~\cite{He:2015eua}. This perspective will be reported elsewhere. 

%\section{Acknowledgements }
%%%%%%%%%%%%%%%%%%%%%%%%%%%%%%%%%%%%%%%%%%%%%%%%%%%%%%%%%%%%%%%%%%%%%%%%%%%%%%%%%%%%%%%%%%%%%%%%%%%
The author thanks Masayasu Harada for discussions.

\end{document}